\documentclass[preprint,journal]{vgtc}       

\pdfoutput=1\relax                   
\usepackage{graphicx}                
\usepackage[aboveskip=1pt, belowskip=1pt]{subcaption}

\usepackage{microtype}                 
\usepackage[font=small,labelfont=bf, skip=0pt]{caption}

\PassOptionsToPackage{warn}{textcomp}  
\usepackage{textcomp}                  
\usepackage{pgf-pie}  
\usepackage{comment}

\usepackage{mathptmx}                  
\usepackage{times}                     
\usepackage{cite}                      
\usepackage{tabu}                      
\usepackage{booktabs} 
\usepackage[nolist]{acronym}
\usepackage{hyperref}

\newacro{vr}[VR]{Virtual Reality}
\newacro{ar}[AR]{Augmented Reality}
\newacro{hmd}[HMD]{Head-Mounted Display}
\newacro{ems}[EMS]{Electrical Muscle Stimulation}
\newacro{imu}[IMU]{Inertial Measurement Unit}
\newacro{bci}[BCI]{Brain-Computer Interface}
\newacro{vwg}[VWG]{Virtual World Generator}
\newacro{moo}[MOO]{Multi-Objective Optimization}
\newacro{dof}[DoF]{Degree of Freedom}
\newacro{ssq}[SSQ]{Simulator Sickness Questionnaire}
\newacro{fov}[FOV]{Field of View}
\newacro{plt}[PLT]{Pareto Least Turns}
\newacro{psp}[PSP]{Pareto Shortest Path}
\newacro{rrt}[RRT]{Rapidly-exploring Random Trees}
\newacro{vims}[VIMS]{Visually Induced Motion Sickness}
\newacro{ddr}[DDR]{Differential Drive Robot}
\newacro{vrise}[VRISE]{VR Induced Symptoms and Effects}

\newcommand{\change}[1]{\textcolor{black}{#1}}

\onlineid{0}

\vgtccategory{Research}
\vgtcpapertype{methodological}

\title{Augmenting Immersive Telepresence Experience with \\ a Virtual Body}

\author{Nikunj Arora, Markku Suomalainen*, Matti Pouke, Evan G. Center, Katherine J. Mimnaugh,\\ Alexis P. Chambers, Sakaria Pouke and Steven M. LaValle }
\authorfooter{
\item
 All authors are with University of Oulu, Finland: firstname.lastname@oulu.fi
 \item * Corresponding author ( markku.suomalainen@oulu.fi )

}

\abstract{We propose augmenting immersive telepresence by adding a virtual body, representing the user's own arm motions, as realized through a head-mounted display and a 360-degree camera. Previous research has shown the effectiveness of having a virtual body in simulated environments; however, research on whether seeing one's own virtual arms increases presence or preference for the user in an immersive telepresence setup is limited. We conducted a study where a host introduced a research lab while participants wore a head-mounted display which allowed them to be telepresent at the host's physical location via a 360-degree camera, either with or without a virtual body. We first conducted a pilot study of 20 participants, followed by a pre-registered 62 participant confirmatory study. Whereas the pilot study showed greater presence and preference when the virtual body was present, the confirmatory study failed to replicate these results, with only behavioral measures suggesting an increase in presence. After analyzing the qualitative data and modeling interactions, we suspect that the quality and style of the virtual arms, and the contrast between animation and video, led to individual differences in reactions to the virtual body which subsequently moderated feelings of presence. %
} 


\teaser{
  \centering
  \includegraphics[width=1.00\linewidth]{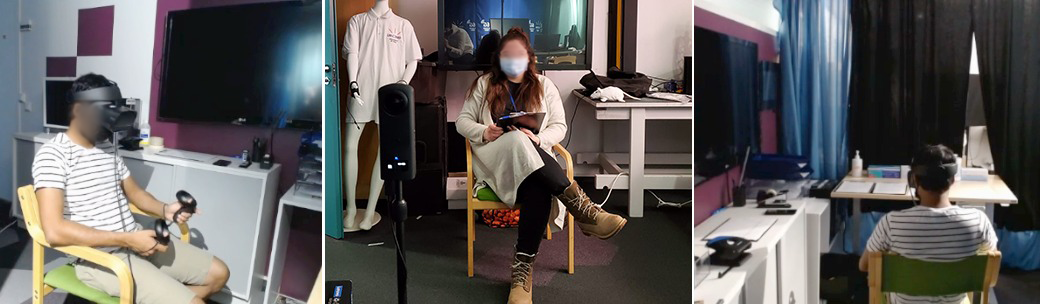}
  \vspace{0.1cm}
  \caption{\change{Head-mounted display telepresence augmented with a virtual body; the participant (side images) is sitting in one place wearing a head-mounted display, but is telepresent in another location through a live streaming camera having a discussion with an experimenter (center image) and can see robot arms in one condition that match their real arms' motions.}}
	\label{fig:teaser}
}

\begin{document}


\firstsection{Introduction}

\maketitle










Marvin Minsky coined the word `telepresence' in his influential 1980s essay, suggesting a new paradigm of remote work where one could carry out complex physical tasks in remote locations while simultaneously receiving rich sensory feedback \cite{minsky1980telepresence}. 
In his view, the biggest challenge of telepresence was achieving the feeling of \textit{presence}, the sensation of `being there'. This remains a challenge in current commercial telepresence robots using regular 2-D screens \change{due to the screens' insufficient immersion. Though robot mobility increases presence over stationary video calls \cite{rae2014bodies}, users still do not really feel as `being there', causing issues such as remote participants speaking less in group work tasks, finding the tasks more difficult \cite{stoll2018wait}, or having discussions less interactive \cite{tree2021psychological} than in-person discussions}. Modern \acp{hmd} and \change{360-degree video cameras can supply the immersion capable of remedying this deficiency and provide users a more similar experience as being physically at the camera's location; adding a panoramic camera on a mobile robot could further increase this feeling. There are a wide variety of use cases for this setup;} for example, people with mobility issues or living too far can join family events, and business people can join events requiring mobility, such as facility tours or workshops, while feeling as if they were really present at the event. 


Following telepresence research, the term `presence' was adopted by the Virtual Reality (VR) research community to describe one of the fundamental illusions elicited by VR systems. While an exact definition of presence is still lacking, most VR scholars agree on presence roughly referring to the sensation of `being there,' similarly to Minsky's original essay (for further discussion, see \cite{skarbez2017survey}). One of the most influential definitions regarding this concept is Mel Slater's framework of immersion and presence, in which the former is defined as the physical system's capability of providing immersive experiences, whereas the latter refers to the subjective feeling of being in a virtual location \cite{slater1997framework}. Later, to reduce confusion amongst terminology, Slater proposed the terms Place Illusion (PI), to specifically refer to the sensation of being in another location \cite{slater_psi:2009}, and Plausibility Illusion (PSI) to refer to the illusion of the virtual scenario actually happening to the user and the general sensation of realism \cite{slater_psi:2009}. Skarbez proposed a framework where the entirety of `presence' would consist of PI, PSI as well as co- and social presence illusions, the latter referring to VR-generated illusions of being in the company of other people or interacting with others, respectively \cite{skarbez2017survey, biocca2001networked}.
\footnote{\change{Similarly, in this article, we also use `presence' to address multiple components, while referring to the subcomponents as PI, PSI and co-presence.}} 


PI and PSI make VR interesting because they enable \textit{realistic responses} in users. For example, a VR user can feel genuine fear of a virtual pit despite having certain knowledge that the pit is not actually there. It has been argued that these illusions and their capability to produce realistic responses are what make VR useful in a number of applications, ranging from therapy to rehabilitation as well as social interaction \cite{slater_psi:2009}. The same property can help achieve a real breakthrough in telepresence robotics. Though  Minsky's original work discussed issues with presence, there are limited studies about presence in telepresence, even though the concept is acknowledged in many ways in robotic telepresence research (for example, \cite{rae2014bodies,stoll2018wait, tree2021psychological, tanaka2014robot,suomalainen2022unwinding}). Thus, research on increasing presence in telepresence is timely and needed.  

The role of proprioception in presence was identified early in VR research; the better the system can accommodate users' physical actions (like render matching multimodal output when the user is moving around), the more immersive the experience, and therefore the greater the capability of eliciting PI \cite{slater1997framework, sanchez2005presence, slater_psi:2009}. The aforementioned requirements include a visible body that matches the user's own proprioception; 
HMD-based VR systems 
achieve this by a virtual body coupled with a tracking system. PSI, on the other hand, is suggested to be dependent o\change{n} \textit{coherence}, which can be roughly defined as the general credibility of the virtual scenario and its ability to meet expectations \cite{skarbez2020immersion}. There are studies, however, that indicate the benefits of having a visible body for PSI as well \cite{slater2010simulating, skarbez:2017}.

Although the role of a virtual body in presence has been investigated in abundance in the field of VR, its role in 360-video, or robotic telepresence, is yet underexplored. We, however, argue that even a virtual nonphysical body augmented on top of a video feed has the potential to enhance the experience of a remote participant using a 360-video based telepresence system. In this paper we present the results of our study concerning the augmentation of a 360-video based telepresence experience with a virtual body and its effects on presence.  





This paper contributes results from a rigorous user study on the effect of having a virtual body in HMD telepresence. After an initial 20-participant pilot, we hypothesized that having a virtual body would increase PI, co-presence, and preference of the user, measured by both questionnaires and behavioral metrics. After estimating effect sizes using the pilot study, we performed a 62-person confirmatory study. However, the confirmatory study did not confirm any of our questionnaire-based hypotheses, and from behavioral metrics we obtained mixed results depending on the exact metric used. We looked into open-ended questions to characterize preferences that could explain the unexpected results, and discovered that individual sentiment towards the robot body, often arising from the individual's perception of the naturalness of the robot body, moderated participants' sense of presence to a large degree. \change{We believe this study, although not confirming our initial hypotheses, will be an important springboard for further research in understanding how individual differences can contribute to the adoption of HMD telepresence technology going forward.}  

\section{Related Work}
Telepresece has different or complementary use cases compared to the common multi-user Collaborative Virtual Environments (CVEs), the earliest of which are already several decades old \cite{noma1995multi}. Contemporary CVEs, such as Rec Room, Glue, or Mozilla Hubs are available for the general public and allow remote participants to communicate in computer-generated virtual space, either through \acp{hmd} or traditional displays. The drawback of CVEs is that they require everyone to participate virtually; a mix of physical and virtual participants is not possible or very complex \cite{koskela2018avatarex}.
In contrast, there are often cases where one or several participants would like to join a physical event, such as a factory tour or grandchild's birthday, remotely. For these types of events, it is important that the remote person also feels present, which is where telepresence robots are useful.

Whereas there is research on using 360-degree cameras as streaming devices during the COVID-19 pandemic to stream, for example, a funeral \cite{uriu2021generating}, there is a limited amount of research on using an \ac{hmd} for face-to-face communication between telepresent and local people. Much work regarding an \ac{hmd} in communication instead focuses on shared experiences, such as an \ac{hmd} user ``riding along" with another  \cite{kasahara2015jackin,tang2017collaboration}. While most publications regarding telepresence robots consider 2-D telepresence robots which stream to a regular screen, there is an increasing amount of recent work regarding how a mobile telepresence robot could stream to an \ac{hmd} to increase the feeling of presence of the remote participant \cite{zhang2020people,jones2021belonging,becerra2020human,mimnaugh2021analysis,suomalainen2021comfort}. The work presented in this paper is geared towards telepresence robots, but can be applied to stationary cameras as well.

\subsection{Presence and Virtual Body}
\label{sec:related_presence}
Over the years, a multitude of studies have been conducted regarding various aspects and their effect on presence in VR. Some examples include locomotion techniques \cite{stavar:2011, slater1995taking}, visual fidelity \cite{zimmons2003influence, slater2009visual, yu2012visual}, passive and active haptic feedback \cite{kim:2017, meehan2002physiological}, and audio \cite{baldis:2001}. The role of proprioception and virtual body for presence in VR is well-known \cite{sanchez2005presence, slater_psi:2009}. The more users can utilize natural body movements to control the VR system, the better \cite{slater1998influence}. Also, having a visible virtual body corresponding to user movements is beneficial for PI \cite{slater1993representations}. This is unsurprising, since \textit{sensorimotor contingencies} \cite{noe2004action}, the match between your own sensorimotor actions and VR output, or \textit{immersion}, is known to be a prerequisite for PI \cite{slater_psi:2009}. Interestingly, however, even though PSI does not depend on sensorimotor contingencies, virtual bodies seem to affect not only PI, but PSI as well \cite{slater_psi:2009}. A study aimed at separating various PI and PSI related aspects found that having a virtual body was important for both PI and PSI \cite{slater2010simulating}. Also, in Skarbez's experiment examining the relative importance of various factors for eliciting PSI, having a virtual body was found to be the most important \cite{skarbez:2017}. 

In the context of VR and telepresence, inhabiting a foreign, or virtual, body is often referred as the \textit{sense of embodiment} (SOE) \cite{kilteni2012sense}. SOE has been defined to consist of the following three subcomponents: \textit{agency} (I control the body), \textit{body ownership} (this is my own body), and \textit{sense of self-location} (my location in relation to the body, distinct from PI) \cite{kilteni2012sense}. There are multiple studies showing that a personalized avatar (or a personalized virtual limb) can increase body ownership \cite{argelaguet2016role,jung2017realme,gorisse2019robot}. However, Lugrin et al.~\cite{lugrin:2015} found that there is a risk of uncanny valley with human-like avatars, and that cartoon-like and robot-like avatars provided better body ownership. Among many other things, SOE has been utilized to induce gender-swapping and race-swapping illusions \cite{bolt2021effects, banakou2016virtual}, and even out-of-body experiences \cite{slater2010first}. 



\subsection{Co-presence}
Co-presence and social presence are presence-related constructs referring to the VR illusions of being together with company. While the exact terminology somewhat differs throughout the literature, Biocca \cite{biocca2001networked} and Skarbez \cite{skarbez2017survey} suggest that the co-presence illusion refers to the illusion of other humans being present, whereas the social presence illusion refers to the illusion of interacting with someone. Co-presence has been the focus of various studies concerning both reactions to virtual characters \cite{garau2005responses}, as well as human-controlled avatars in CVEs \cite{casanueva2001effects,koskela2018avatarex}. 

The role of virtual bodies in co-presence and social presence illusions has been studied in the context of both CVEs and robotic telepresence. In a purely virtual setting, a realistic hand created the best social presence when communicating with sign language \cite{yoon2020evaluating}. Tanaka et al.  \cite{tanaka2014robot} found embodying a physical robot enhanced social telepresence in remote face-to-face communication. There exist some studies where video footage has been combined with virtual elements; in a study similar to ours, there was no difference in social presence between realistic and cartoon-like avatars in a collaboration between AR and VR users \cite{yoon2019effect}. In a study focusing mostly on user experience, Danieau et al.~\cite{danieau2017enabling} did not find significant differences for PI when augmenting pre-recorded 360 video with various virtual bodies; \change{however, Chen et al.~\cite{chen2017effect} did notice an increase in presence when users could see their hands and a table in front of them. In this study, the difference to the previous ones is that the discussion was live, not pre-recorded, forcing the participants to engage in conversation as well}.

To conclude, there is interest and need for our study, namely in understanding the importance of seeing your ``own" virtual body in VR-based telepresence. Moreover, we envision that future telepresence robots might have physical robotic arms for manipulation tasks and to facilitate communication. Since we were not interested in researching individual virtual bodies, we decided to display the user's body to themselves as robot arms and a robot torso instead of showing realistic looking human arms or any visualization of legs or the lower body. This also afforded us the benefit of avoiding any uncanny valley effects.

\section{Hypotheses}
We performed an initial pilot study to further specify our hypotheses and analysis methods. Based on related literature, we predicted that adding a body would make the study participants feel greater presence, and as well that they would prefer having a body. We measured PI both with questionnaires and behavioral metrics, whereas co-presence and preference are measured with questionnaires only. Based on the results of the pilot study, we tested the following hypotheses:
\vspace{-0.1cm}
\begin{itemize}
    \itemsep=0em
    \item H1: Having a representation of an artificial robot body (as opposed to no virtual body at all) in a telepresence situation increases the feeling of PI
    \item H2: Having a representation of an artificial robot body (as opposed to no virtual body at all) in a telepresence situation increases the feeling of co-presence
    \item H3: The users prefer having the virtual body
\end{itemize}

\section{Methodology}

\subsection{Participants}
Participants gave written informed consent in accordance with the local ERB prior to participating in the study. Our required sample size was derived from our pilot study of 20 participants where the smallest effect size of interest was the \change{Slater-Usoh-Steed} (SUS) questionnaire score difference between body and without body conditions (\textit{dz}=0.35; see Section \ref{sec:analysis}). We performed \textit{a priori} power analysis using G*Power software and determined that 62 participants would be needed to achieve 80\% power to detect an effect of this magnitude or greater while still obtaining an equal number of participants per group for the between-participants hand motion comparison. Our final 62 participant sample had a mean age of 30.1 years old (\textit{sd}=6.33) and consisted of 20 females, 41 males, and one participant who preferred to self-describe their gender.

When replacing participants after initial exclusions, by error, replacement participants were not distributed into conditions in the same manner as the excluded participants they replaced, resulting in 33 participants in the “body first” condition order and 29 participants in the “without body first” condition order. This created an imbalanced ratio of 33 and 29 participants per group in the between-participants analyses (hand motion) and 62 participants per condition in the within-participants analyses (all others).

\subsection{Exclusion Criteria}
Preregistered exclusion criteria stated that participants who did not complete the entire experiment’s protocol, or for whom the onset of the rat throw event (see Section \ref{sec:procedure}) was not visible in the recording frame, would be excluded from the experiment. Protocol completion and recordings were verified after data from 62 participants had been collected, and 12 new participants were recruited to replace participants excluded due to a camera recording malfunction.


\subsection{Study Setup}
To test the effects of having a virtual body in telepresence, we presented participants with a scenario where they are introduced to equipment in a research laboratory. Participants were seated and stationary while one of the authors assumed the role of the laboratory introduction host and followed a predetermined script to interact with every participant. This researcher shall henceforth be referred to as the host. 

The script is broken into two parts, where one session is experienced with a virtual body and the other without a visible body, \change{in alternate orders counterbalanced among participants to make sure each session was experienced both with and without a body}. 
Incoming participants were further counterbalanced by gender such that there would be an equal number of males and females experiencing each condition order \change{to preclude any potential effects, though we did not expect to see them.} 
The script was designed in such a way that the host did not need to rely on the participants' responses to progress, though responses were solicited to make the participants more engaged. Throughout the interaction, the participant is asked to indicate the location of various objects in the lab so as to make the conversation interactive and to give the participant opportunities to use their hands (even if participants could not actually 'touch' anything they see). We note that we did not tell the participants whether the host could see their hands and they were never explicitly asked to use their hands; this was a deliberate decision, as we focused on how the body affected the participants themselves and wanted to observe the emergent participant behavior. \change{Unbeknownst to the participant, the curtain separating the participant and host was slightly parted immediately prior to the beginning of the interaction so that the host could actually see the participant's real hands regardless of the participant having the virtual body or not. However, the host was at all times blind in regards to whether the participant was experiencing the with or without body condition. The curtain was lowered again before the participant took off the \ac{hmd} to make sure they did not realize the interaction occurred so close physically.}  

\begin{figure}[b]
 \vspace{-0.8cm}
 \centering
 \includegraphics[width=\columnwidth]{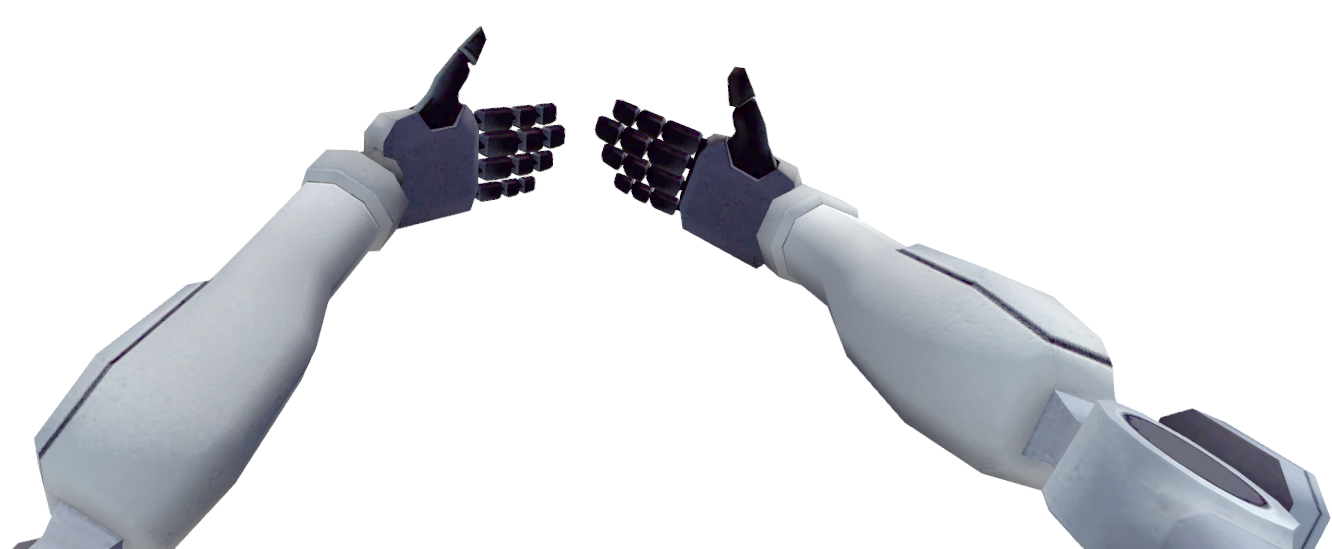}
 \caption{\change{The user's view of the robot arms with the background removed.}}
 \label{fig:hands}
 \vspace{-0.5cm}
\end{figure}

The participant's arms were shown as robot arms \change{(seen in Fig.~\ref{fig:hands})} for two reasons; first, we wanted to investigate through simulation whether a future telepresence robot having real, physical arms could increase the experience also for the remote user. Second, even though personalized avatars have shown stronger presence and agency \change{\cite{jung2017realme}}, other results indicate that cartoon or robot-like arms can be a good substitute \change{\cite{lugrin:2015}}. We wanted to avoid ending up in the uncanny valley with a human-like hand \cite{lugrin:2015}, and personalizing arms to make them look like the user's own arms was not a motivation for this study. Inverse kinematics tracking for arms and simple finger gestures were provided by Oculus Rift S controllers.  
No other body parts were tracked; foot tracking has been shown to have no significant effect on the overall result  \cite{pan:2019} unless they analyzed the participants' movement behaviour; as the participants didn't locomote in our study due to the camera (in this study Ricoh Theta Z1 \change{using full HD resolution}) also being stationary, the body's legs were hidden and only the torso and arms were visible. \change{The robot arms were based on the Space Robot Kyle from the Unity Asset Store}\footnote{\url{https://assetstore.unity.com/packages/3d/characters/robots/space-robot-kyle-4696}} \change{scaled up 1.3 times on all axes. The arm model had similar joints as a human, and the total length of each arm was 0.8 meters.}

Keskinen et al. \cite{keskinen:2019} observed that the user's own height with respect to the camera height had no effect on the viewing experience and concluded that a height of 150 cm was suitable for the camera mount, irrespective of the user's actual height or posture. However, during the initial tests of the system, users felt too high at 150 cm while sitting. Ultimately for both the pilot and main experiment, the camera height was set to be equal to the eye height of the host when seated (117 cm). 

All connections were wired to minimize latency. Thus, the camera was chosen to be placed in close proximity to the participant's actual location, both to reduce latency in the video feed and provide directional audio (shown to be important for creating realistic feelings in a remote environment \cite{baldis:2001}) to the participant without relying on a 3D audio technique. \change{The camera was connected via USB cable to the same laptop that the VR headset was connected to. 
The 360-degree camera footage was accessed as standard webcam footage in Windows from Unity, where the arms were added and the combination then streamed to the HMD.} In practice, the participant was sitting only several meters from the camera and the host, but the participant could not see the host setting due to a separating curtain. The host always entered and exited through a different door than the one used by the participant to avoid the participant realizing they were telepresent only behind a curtain. 
Possible lip synchronization issues, due to small delay in the video but no delay in the audio, were prevented by the host wearing a mask due to the COVID-19 protocols.

\subsection{Procedure}
\label{sec:procedure}
Upon arrival, participants were greeted by another researcher (the experimenter) and asked to sit in a chair that was fixed at a predetermined location. They were then asked for consent, after which the experimenter read out the instructions, presented the controls, and told the participant how to put on the VR headset. When they were ready to begin, the experimenter covertly signalled the host to begin the interaction. After the host entered the room, the curtains which separated the participant from the location of the camera were moved slightly without the participant's knowledge to allow the host to see the participant's gestures, both with and without the virtual body. 

Each participant was instructed to hold the controllers regardless of condition to avoid creating a bias due to not holding the controllers even when no virtual body was displayed. 
Recordings of participants were used to monitor their reactions to events in the script. Aside from video data, the position coordinates and Euler angles of the HMD and both controllers in Unity's world coordinate system were also recorded for every frame. \change{During the pilot study, we had an induction phase where the participants were told to practice different finger gestures using the VR controllers (similarly as in, for example, \cite{ebrahimi2018investigating}). However, this caused  the participants to focus on memorizing the gestures and led to the expectation that they would need to use the optional gestures; thus, the induction phase did not evoke the emergent behavior that we wanted to see. Hence, we modified the instructions to reduce emphasis on practicing the gestures for the subsequent confirmation study.}

In an attempt to elicit movement from the participant and gauge an implicit sense of presence, one key event took place during each session. \change{In the first session, a plush rat doll was thrown towards the camera. After the participant was asked if they could see the rat, the host picked it up to show them and then shouted that the rat bit her, throwing the rat at the camera in the process}. In the second session, the host walked close to the camera to pick up the thrown item, which could be interpreted by participants as an invasion of personal space. We chose hand motion as our dependent variable in the rat throw event and head motion as our dependent variable in the approach event based on the types of motion elicited in  pre-pilot tests. After the host left, the curtains were covertly drawn closed again before asking the participant to remove the headset. We chose to focus only on hand motion in response to the rat throw event after reviewing pilot data (see Section 5.1). Depending on condition order, some participants had a body during the rat throw event while others did not, requiring a between-participants analysis of hand motion within the larger within-participants design (see Section 4.7).

After each session, participants were asked to fill out a questionnaire regarding their experience. After both sessions were completed, they were presented with a post-experiment questionnaire that queried demographic information and overall opinions. \change{Finally, the participants received 20€ Amazon vouchers for participating in the study.}

\subsection{\change{Measures}}

Three sets of questionnaires were presented to the participant; two after finishing each of the sessions, and one post-experiment questionnaire after both sessions are over and their respective questionnaires have been submitted. The Google forms service was used to design and present the questionnaires to the participants. The quantitative questions were designed using a Likert scale \cite{mcleod:2019} with a range of 1 to 7.

The questionnaires after the sessions were divided into three or four sections, depending on the virtual body condition for that session. The first section determines the participant's sense of PI. For this component, we used the extended version of the SUS questionnaire \cite{slater1994depth, usoh:2000}, using \textit{``remote environment"} and \textit{``sitting with the host"} in place of virtual environment (see Appendix 1). \change{We believed that the SUS would be the most appropriate presence questionnaire for this part of the study because the SUS focuses solely on PI.}




The second section of the questionnaire was used for measuring the co-presence with the host. The questionnaire was adapted and modified based on the co-presence questionnaire presented by Casanueva et al. \cite{casanueva:2001} and the co-presence subsection of the social presence questionnaire by Biocca et al. \cite{biocca2001networked}. The exact questions were chosen and modified to fit the scope of this study (see Appendix 1).


The third section of the questionnaire collected qualitative data regarding Breaks In Presence (BIP) \cite{garau2008temporal}. 
The final section was exclusive to the condition in which the participant had a visible body. This questionnaire (a subset of questions taken from Gonzalez-Franco and Peck embodiment questionnaire \cite{gonzalez-franco:2018}) measures participant's sense of embodiment to the virtual avatar. 



After all these questionnaires were filled in both sessions, the participants were presented with a post-experiment questionnaire; forced-choice questions on which condition they preferred and which condition induced more co-presence, and their demographic information such as gender, age, video gaming history and history with VR. This final, more overt set of questions was reserved until all conditions had been completed in order to avoid the influence of demand characteristics on the previous, more covert sets of questionnaires.

\subsection{Pilot Study} 

We conducted a separate pilot study of \change{20 participants (13 male; mean age of 28.5, \textit{sd}=4.52)} to fine tune protocol and estimate effect sizes for each analysis. The procedure and analyses to be used in the main study were then preregistered at \url{https://osf.io/daks8}. Protocol between the pilot and confirmatory study differed in that the host asked and repeated back the participant’s name at the beginning of the interaction in the main study, rather than beginning the interaction without introductions as in the pilot, in order to communicate that the experience was interactive rather than a passive recording. \change{Also, participants were not required to practice gestures with the controller before the study, though it was suggested that they try,} and one request to ``point" to an item was replaced with asking ``do you see" the item.

\subsection{Analyses}
\label{sec:analysis}
Confirmatory analyses were selected based on pilot data. All other analyses are exploratory and are described under Section 5.3. Confirmatory analyses were preregistered as one-tailed tests, when applicable, while exploratory analyses were always two-sided tests, when applicable. We tested five preregistered hypotheses, applying the Holm correction to correct for multiple comparisons. Applying the Holm correction for multiple comparisons gave the following critical \textit{p}-values for our five analyses: .01, .0125, .017, .025, and .05. After obtaining observed p-values for each confirmatory analysis, these thresholds applied to tests of hand motion, preference for body condition, explicit evaluation of PI, co-presence scores, and SUS scores, respectively.  

We sought to measure PI implicitly, via behavior, explicitly, via asking participants directly in which session they felt more present, and somewhere between these two extremes, via the SUS questionnaire. Co-presence and preference were measured only by questionnaire or by asking participants directly whether they felt having a body improved the overall experience, respectively. Given that directly asking participants about their sense of PI, or for their preference, could induce demand characteristics, we place less weight on the results obtained through this method but report them nonetheless for completeness.


SUS score was calculated as the number of SUS questionnaire items given a 6 or 7 rating by a participant, resulting in a score that could range from 0 to 6 for each participant \cite{slater1994depth}. We performed a Wilcoxon signed-rank test (one-sided) to test for greater SUS scores in the body condition than in the without body condition. 

Hand motion in response to the rat throw event was analyzed in the following way: an experimenter identified the frame (framerate: 60 fps) at which the rat left the host's hand and marked this frame as the onset of the event for each participant. The position of the participant's hands, as marked by Oculus Rift S controllers, at the time of this onset frame served as the initial starting position for motion quantification. Motion was quantified by integrating the total distance in meters traveled from this initial starting position, in three dimensions, for each hand, over the next 200 frames. The integrations for left and right hands were summed to obtain a single measure of motion for each participant. As robustness tests, we also calculated motion as the average distance over the response window of summed left and right hands from a 10 frame pre-window motion baseline, as well as the maximum distance traveled over the response window of summed left and right hands from the same 10 frame pre-window motion baseline. We performed a Shapiro-Wilk test and found that the motion data was non-normally distributed, \textit{W}=.76, \textit{p}$<.001$. We thus tested for greater hand motion in the body group than in the without body group using a non-parametric Wilcoxon rank-sum test (one-sided).

Explicit evaluation of PI (indication of greater PI in the body condition than in the without body condition) was analyzed using an exact binomial test (one-sided). 

Co-presence scores were calculated similarly to SUS scores as the number of items in the co-presence questionnaire given a 6 or 7 rating by a participant. Because our co-presence questionnaire only had four items in comparison to the six of the SUS questionnaire, co-presence scores could range from 0 to 4 for each participant. We performed a Wilcoxon signed-rank test (one-sided) to test for greater co-presence scores in the body condition than in the without body condition.

Preference for body condition (preference for the body condition over the without body condition) was analyzed using an exact binomial test (one-sided). 

\section{Results}
\subsection{Pilot Results}
\label{sec:pilot}

\change{The pilot study consisted of 20 participants using a within-participants design. All analyses were run within-participants except head and hand motions, which were between-participants and thus had only 10 people per group)}. Pilot results indicated benefits of having a virtual body in regard to PI, co-presence, and preference. We observed a strong influence of having a virtual body each in terms of resulting hand motions in response to the rat throw event, co-presence scores, explicit presence, and explicit preference, while the effects in terms of PI as measured by SUS scores were moderate. In contrast, the virtual body manipulation produced negligible effects on resulting head motions in the approach event.

SUS scores in the body condition (\textit{m}=3.80) were not statistically greater than SUS scores in the without body condition (\textit{m}=3.05) as indicated by a Wilcoxon signed-rank test (two-sided), \textit{Z}=1.48, \textit{p}=.14, \textit{r}=0.33, suggesting there was no substantial increase in PI when participants had a virtual body in this sample, as measured by the SUS survey. 

The distribution of hand motion was non-normal as indicated by a Shapiro-Wilk normality test, \textit{W}=.85, \textit{p}=.006, thus a non-parametric test was used. During the rat throw event, motion was significantly greater in the body group (\textit{m}=1.30) than in the without body group (\textit{m}=0.41) as indicated by a Wilcoxon rank-sum test (two-sided), \textit{Z}=2.53, \textit{p}=.012, \textit{r}=0.57, suggesting that the inclusion of a virtual body increased participants' implicit sense of PI during the rat throw event. 

The distribution of head motion was non-normal as indicated by a Shapiro-Wilk normality test, \textit{W}=.89, \textit{p}=.025, thus a non-parametric test was used. During approach event, motion was not significantly greater in the body group (\textit{m}=0.35) than in the without body group (\textit{m}=0.25) as indicated by a Wilcoxon rank-sum test (two-sided), \textit{Z}=0.04, \textit{p}=.97, \textit{r}=0.01, suggesting that the inclusion of a virtual body had little effect on participants’ implicit sense of PI when rendered during the approach event.

Due to the small effect size related to the approach event, and the resulting onerous number of participants that would be required to reliably detect such an effect given it truly exists, the analysis of head motion was not included in our subsequent confirmatory experiment. We initially considered the rat throw and approach events to be equivalent in terms of events that could elicit motion from participants. After reviewing the pilot results, however, we reasoned that the difference between the effects might derive from differences in the time scales and emotional aspects of the events. In particular, the rat throw is a sudden onset, physically threatening event, whereas the approach is more of a medium onset, socially discomforting event. It is also perhaps not surprising that including visible arms would have a larger effect on hand motion than on head motion. 

When asked explicitly which condition lent greater PI, 18 out of 20 participants selected the body condition, and this preference towards the body condition was significant in an exact binomial test (two-sided), $p<.001$, suggesting that the inclusion of a virtual body increased participants’ explicit sense of PI.

Co-presence scores in the body condition (\textit{m}=3.00) were significantly greater than co-presence scores in the without body condition (\textit{m}=2.05) as indicated by a Wilcoxon sign-rank test (two-sided), \textit{Z}=2.45, \textit{p}=.014, \textit{r}=0.55, suggesting there was an increase in co-presence when participants had a virtual body, as measured by the co-presence survey.  

When asked explicitly about their preference for virtual body, 16 out of 20 participants selected the body condition, and this preference towards the body condition was significant in an exact binomial test (two-sided), \textit{p}=.012, suggesting that participants preferred having a virtual body to not having any body representation at all.

\subsection{Confirmatory Results}

The pattern of results in the main study differed from the pattern observed in the pilot. We found some support for an increase in implicit presence provided by the virtual body, but little to no support for benefits in regard to PI, co-presence, explicit presence, or explicit preference upon initial analysis.

\change{\textbf{Having the body did not increase presence as measured by the SUS or when asked directly.}} SUS scores in the body condition (\textit{m}=2.69) did not differ from SUS scores in the without body condition (\textit{m}=2.74) as indicated by a Wilcoxon sign-rank test (one-sided), \textit{Z}=-0.48, \textit{p}=.63, \textit{r}=-0.06, providing no evidence of an increase in PI when participants had a virtual body, as measured by the SUS survey. When asked explicitly which condition created greater PI, 36 out of 62 participants selected the body condition, though this bias towards the body condition was not significant in an exact binomial test (one-sided), \textit{p}=.13. providing no evidence that the inclusion of a virtual body increased participants’ explicit sense of PI.



\begin{figure}%
  \centering
    \begin{subfigure}{0.33\columnwidth}
        \includegraphics[width=\columnwidth]{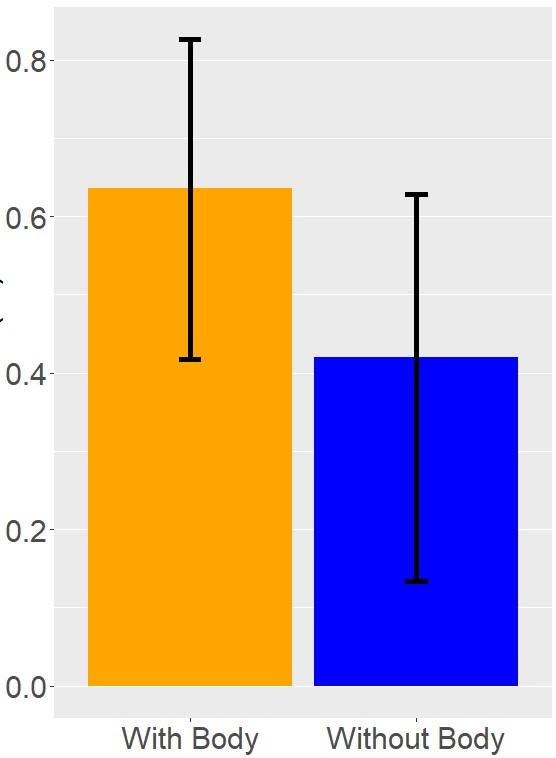}
        \caption{}
    \end{subfigure}\hfill
    \begin{subfigure}{0.33\columnwidth}
        \includegraphics[width=\columnwidth]{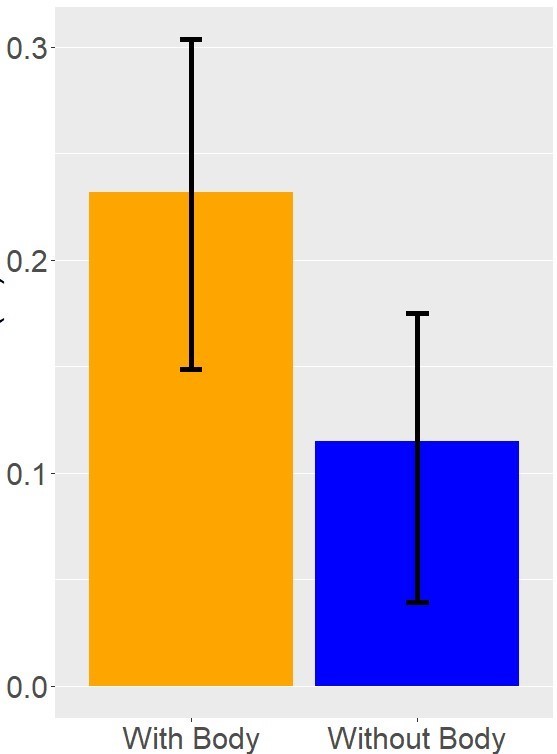}
        \caption{}
    \end{subfigure}\hfill%
    \begin{subfigure}{0.33\columnwidth}
        \includegraphics[width=\columnwidth]{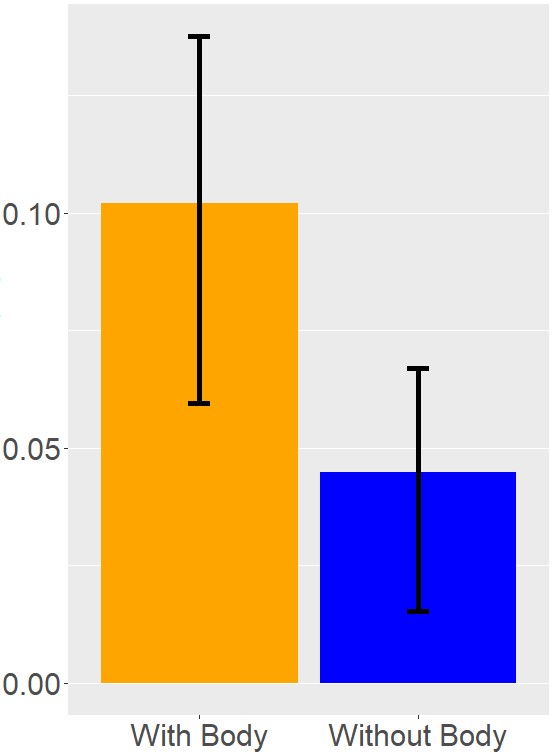}
        \caption{}
    \end{subfigure}\hfill%
    \caption{Hand motion data from the rat throw session, measured with (a) integration, (b) max vector, and (c) mean vector. Motion, in meters, is on the y axis. Error bars represent 95\% bootstrapped confidence intervals.}
    \label{fig:motion_data}
    \vspace{-0.5cm}
\end{figure}

\change{\textbf{There is partial evidence that having the body increases presence through behavioral metrics.}} The distribution of integrated hand motion was non-normal as indicated by a Shapiro-Wilk normality test, \textit{W}=.75, $p<.001$, thus a non-parametric test was used. During the rat throw event, motion was significantly greater in the body group (\textit{m}=0.636) than in the without body group (\textit{m}=0.420) as indicated by a Wilcoxon rank-sum exact test (one-sided), \textit{Z}=2.53, \textit{p}=.011, \textit{r}=0.32, although this difference did not survive the Holm-corrected alpha threshold ($\alpha=.01$). Using the motion average robustness metric rather than motion integration (see Section \ref{sec:analysis}), motion was significantly greater in the body group (\textit{m}=0.102) than in the without body group (\textit{m}=0.045) as indicated by a Wilcoxon rank-sum exact test (two-sided), \textit{Z}=2.92, \textit{p}=.003, \textit{r}=0.37. Additionally, the maximum vector robustness metric (see Section \ref{sec:analysis}) also revealed significantly greater motion in the body group (\textit{m}=0.232) than in the without body group (\textit{m}=0.115) as indicated by a Wilcoxon rank-sum exact test (two-sided), \textit{Z}=2.93, \textit{p}=.003, \textit{r}=0.37. Despite the failure to survive correction for multiple comparisons for the confirmatory study's integrated motion effect, we take the combination of the pilot motion effect, the confirmatory motion effect, and both robustness checks in the confirmatory experiment as some evidence suggesting that having a virtual body offered participants an increased sense of implicit PI on an immediate time scale, given that each effect is statistically significant, or borderline statistically significant, in the predicted direction (see Fig.
~\ref{fig:motion_data}). 

\change{\textbf{Having the body did not increase co-presence.}} Co-presence scores in the body condition (\textit{m}=2.10) were not significantly greater than co-presence scores in the without body condition (\textit{m}=2.02) as indicated by a Wilcoxon signed-rank test (one-sided), \textit{Z}=0.85, \textit{p}=.40, \textit{r}=0.11, providing no evidence of an increase in co-presence when participants had a virtual body, as measured by the co-presence survey.



\change{\textbf{There is, at best, weak evidence that the body was preferred.}} When asked explicitly about their preference for virtual body, 39 out of 62 participants selected the body condition, and this bias towards the body condition was significant in an exact binomial test (one-sided), \textit{p}=.028, however this difference did not surpass the Holm-corrected alpha threshold ($\alpha=.0125$), providing at best weak evidence in combination with the pilot results that participants preferred having a virtual body to not having any body representation at all.

\subsection{Exploratory Results}

\subsubsection{Confound Analysis}
\change{\textbf{We observed no confounds related to the order of sessions and stimuli.}} Here we describe tests for the influence of unintended confounding variables; namely, order effects potentially arising from experiencing the body condition in the first or second session, and session effects potentially arising from differences between the first session containing the ``rat throw" event and the second session containing the ``approach" event. We must note that we are unable to distinguish between an effect of session independent of mere time spent in the HMD, as script constraints dictated that the session containing the rat throw event always preceded the second session containing the approach event. We observed no order effects related to whether the body condition came in the participant's first or second session, as indicated by Wilcoxon rank-sum tests (two-sided) for SUS scores, \textit{Z}=0.42, \textit{p}=.68, \textit{r}=0.05, or co-presence scores, \textit{Z}=1.49, \textit{p}=.14, \textit{r}=0.19. Likewise, we observed no session/HMD accommodation effects, as indicated by Wilcoxon signed-rank tests (two-sided) for SUS scores, \textit{Z}=0.79, \textit{p}=.42, \textit{r}=0.10, or co-presence scores, \textit{Z}=1.91, \textit{p}=.056, \textit{r}=0.24. We also observed no order effects, as indicated by two-sample proportion tests (two-sided) in terms of forced choice PI, $\chi^2$(1,\textit{N}=62)=0.90, \textit{p}=.34, or forced choice preference $\chi^2$(1,\textit{N}=62)=0.43, \textit{p}=.51. Confirmatory motion data was only analyzed from the first session as a between-participants test and thus the confounds described would not apply. 

\subsubsection{Qualitative Data}
\label{sec:qualdata}
We collected open-ended data to gather insights into why participants did or did not prefer having a virtual body. The data we collected included BIPs using open-ended responses, and typed open-ended responses that were justifications to forced-choice questions regarding PI and preference. The open-ended data was analyzed using the thematic analysis method with inductive approach \cite{patton2005qualitative}. In the first stage of analysis, two researchers independently identified codes from the response data and mutually agreed on the codings in the second phase.

\change{\textbf{The stimulus events caused the most breaks in presence, with the \textit{visual proportions} and the \textit{virtual body} gathering the next most mentions.}} The BIP related data contained a relatively small number of remarks regarding the virtual body (4 occurrences), or the lack thereof (6 occurrences). According to questionnaire data, the most common causes for BIPs in both conditions were the \textit{rat throw} (8 and 11 occurrences in no-body and body conditions, respectively)  and \textit{approach} (11 and 9 occurrences respectively), the same events we used to collect behavioral responses. Aside from these events, \textit{visual proportions} (7 occurrences without body, 4 occurrences with body) were a common cause for BIPs (\textit{``when looking down  I felt I am too tall"} and \textit{``When I looked around and realized my proportion was out of place. Host also looked like a giant when they came near me."}). The remaining responses were mostly concerning unrelated disturbances such as external noises, comfort, video quality and HMD related issues. 

\begin{figure}[tb]
 \centering
 \includegraphics[width=0.7\columnwidth]{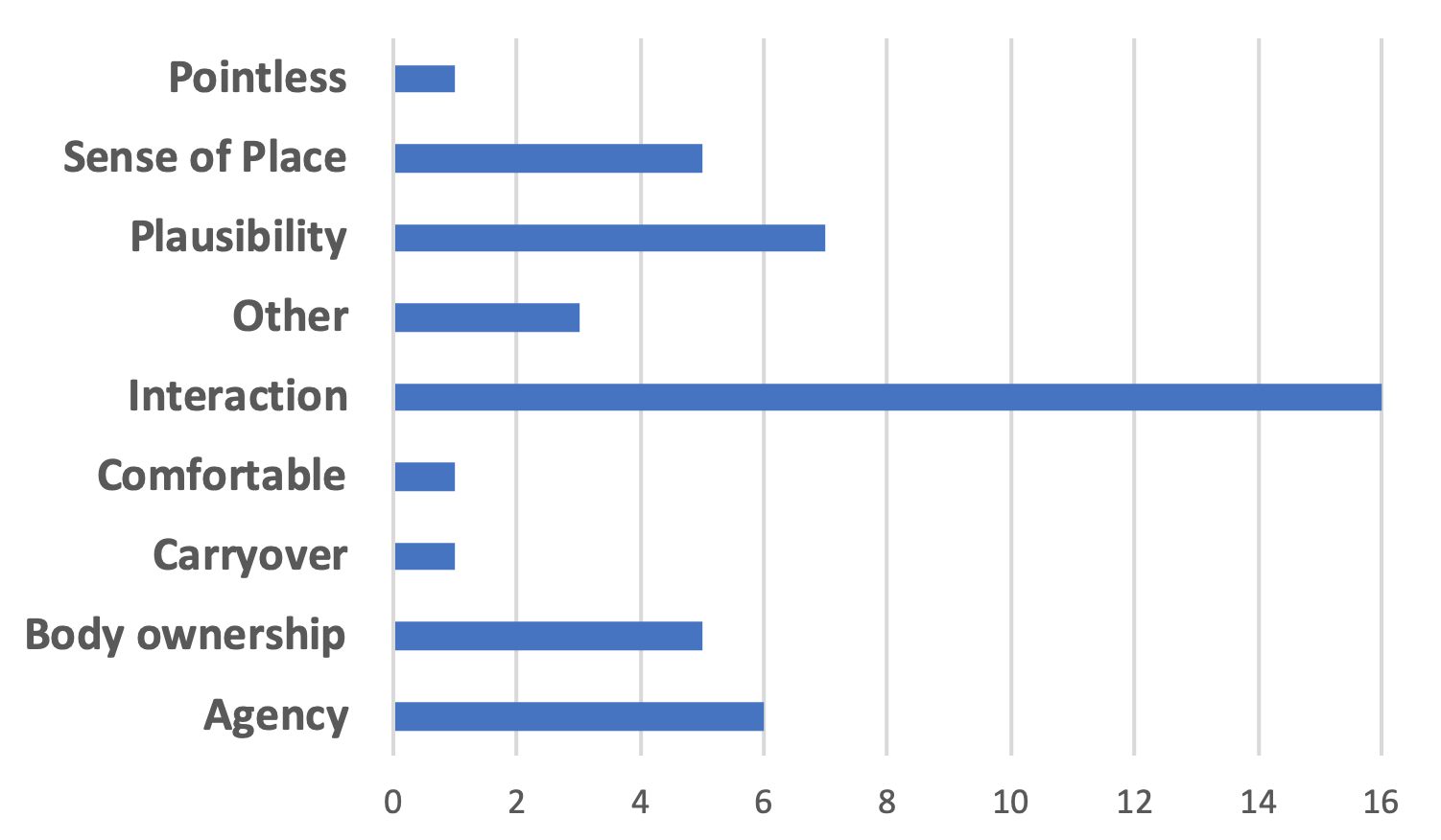}
 \caption{Qualitative codes for participants who considered virtual body beneficial for their PI.}
 \label{fig:PI_w_body}
 \vspace{-0.5cm}
\end{figure}

\change{\textbf{Participants who had a better sense of PI with the body thought the ability to interact was the biggest reason why, followed by answers related to \textit{plausibility} and \textit{agency}.}} Thirty six out of 62 participants considered the virtual body to give a better sense of being next to the guide (PI). For these participants, the three most popular responses concerned \textit{interaction} (16 occurrences), \textit{plausibility} (7 occurrences), and \textit{agency} (6 occurrences). \textit{Interaction} related responses concerned the body being beneficial for either interacting with the guide (\textit{``because i could use my hands to interact with the host"}), or the environment, (\textit{``It gave me a sense of feeling of being able to interact with the environment"}), even though the remote environment did not contain interaction affordances. Responses coded as \textit{Plausibility} contained aspects that could be attributed under PSI or coherence \cite{skarbez2020immersion}. These responses concerned either general sensation of credibility or realism (\textit{``It felt normal to see your hands in front of you"}), or matching expectations (\textit{``less disconnect with what I expect to see. Expecting to see hands but the robot has none causes confusion"}). \textit{Agency} referred to being able to control the virtual body without any specific interaction related context (\textit{``Because there was a sense of physical autonomy"}). All codes  and their frequencies for participants who considered the virtual body enhancing PI can be seen in Fig. \ref{fig:PI_w_body}.

\begin{figure}[tb]
 \centering
 \includegraphics[width=0.7\columnwidth]{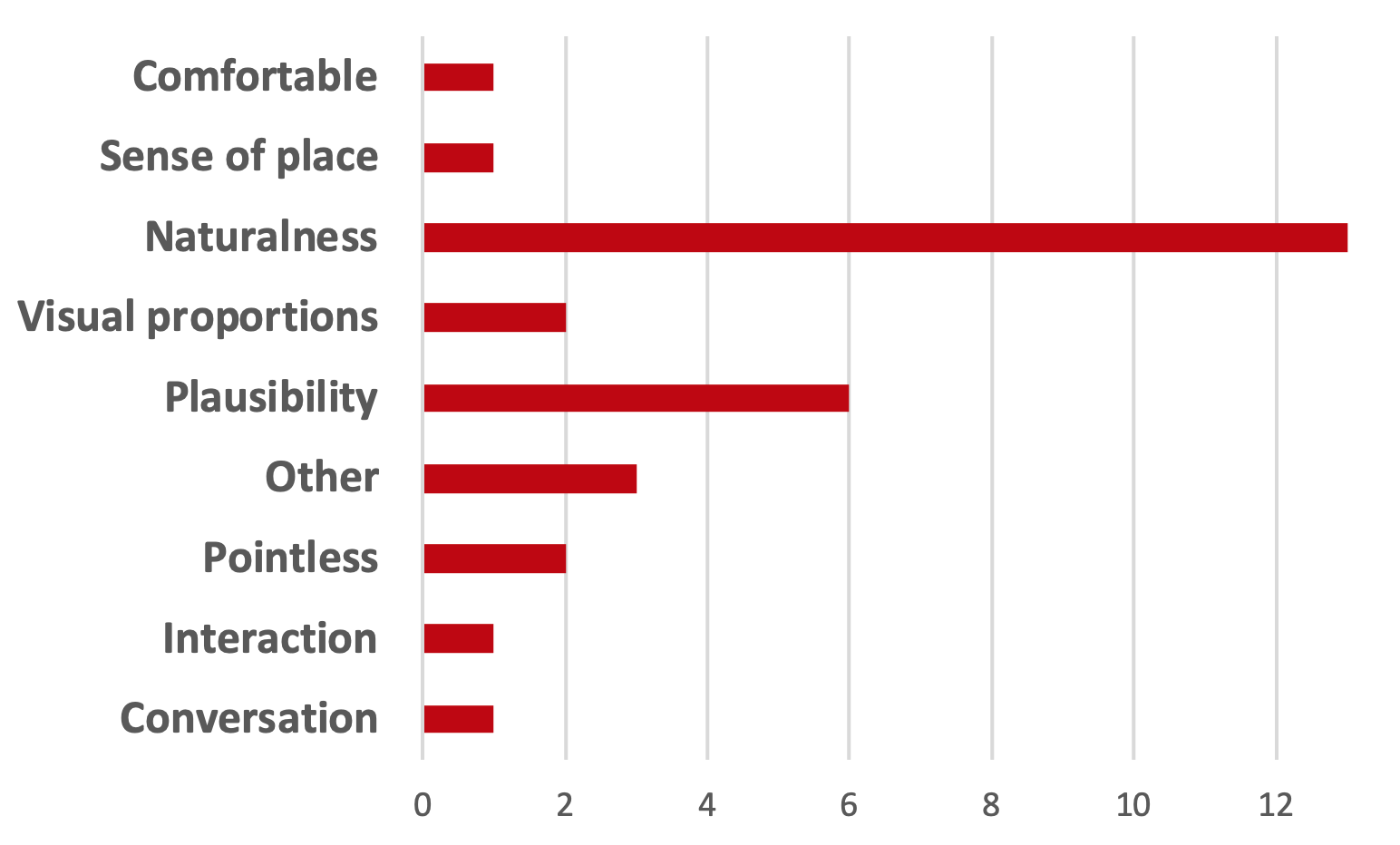}
 \caption{Qualitative codes for participants who did not consider the virtual body as beneficial for their PI.}
 \label{fig:PI_wo_body}
 \vspace{-0.5cm}
\end{figure}

\change{\textbf{Among participants for whom the virtual body did not enhance their PI, the most popular reasons why were \textit{naturalness} (13 occurrences), \textit{plausibility} (6 occurrences) and \textit{other} (3) occurrences}}. Responses coded as \textit{naturalness} contained remarks about the virtual body seeming too artificial or unnatural (\textit{``The robot body wasn't very natural"}). Examples of \textit{plausibility} related responses include \textit{``Somehow seeing the robot hands broke the illusion of actually being somewhere else, and made it feel more like a video game. I immediately did not feel as immersed as in the first time."} and \textit{``Robot  body does not felt [sic] like I am living in reality"}. The category \textit{other} contained responses that did not give a particular reason, were difficult to interpret or did not fit other existing categories (for example, \textit{``i am not sure. i felt the first one was more clear to visualize"}). All codes  and their frequencies for participants who did not consider the virtual body enhancing PI can be seen in Fig. \ref{fig:PI_wo_body}.

\begin{figure}[t]
 \centering
 \includegraphics[width=0.7\columnwidth]{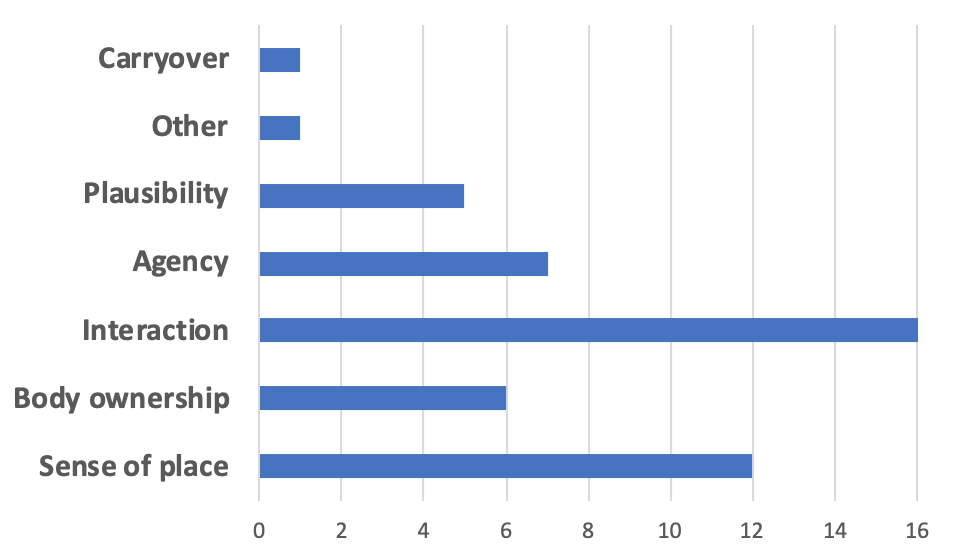}
 \caption{Qualitative codes for participants who considered the virtual body as generally improving their experience.}
 \label{fig:Experience_yes}
 \vspace{-0.5cm}
\end{figure}

\change{\textbf{People who preferred the virtual body named \textit{interaction} and \textit{sense of place} as the biggest factors.}} Thirty nine out of 62 participants considered the virtual body to improve the overall experience. Again, \textit{interaction} (16 occurrences) related responses were most popular (e.g., \textit{``Because I could interact more with the host"}). The second most popular category was \textit{sense of place} (12 occurrences) referring to PI (\textit{``Seeing the robot move in the virtual environment made me feel like I was there. Other way around I felt like I was just watching the room via a camera."} and \textit{``It gave me the feel that I am in some other environment"}). Again, \textit{agency} (7 occurrences) was the third most popular category (\textit{``I have feedback on my senses eg Hand movement"}). In Fig. \ref{fig:Experience_yes} are codes and their frequencies for participants considering the virtual body to improve their experience.

\begin{figure}[t]
 \centering
 \includegraphics[width=0.7\columnwidth]{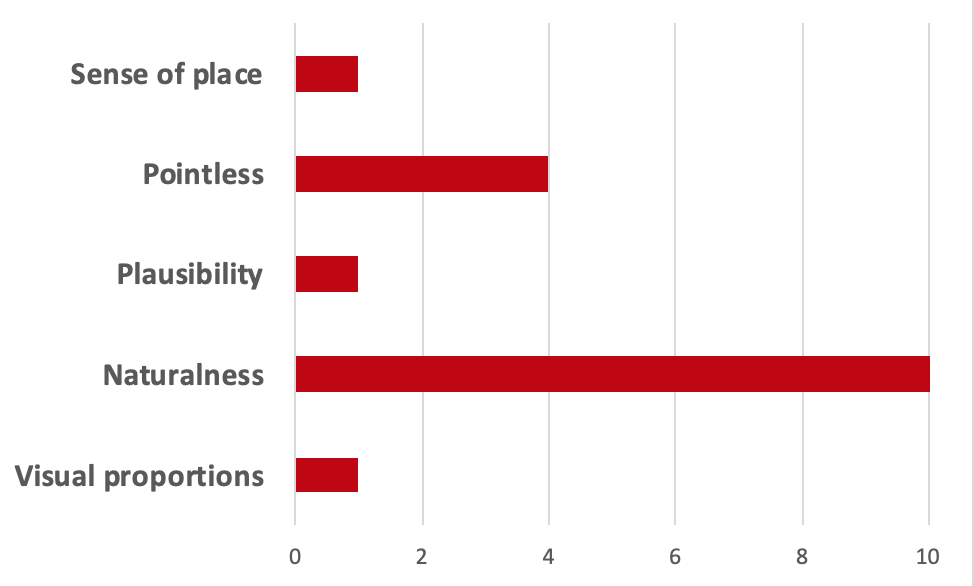}
 \caption{Qualitative codes for participants who did not consider the virtual body as generally improving their experience.}
 \label{fig:Experience_no}
 \vspace{-0.5cm}
\end{figure}

\change{\textbf{For participants who disliked the virtual body in terms of overall experience, the most popular responses why were related to \textit{naturalness} (10 occurrences) and \textit{pointlessness} (4 occurrences)}} (for example, \textit{``Because it had no physical function"}). The remaining codes \textit{sense of place}, \textit{plausibility}, and \textit{visual proportions} were each mentioned once. All codes and their frequencies for participants who did not consider the virtual body improving their experience can be seen in Fig. \ref{fig:Experience_no}.

\subsubsection{Modeling Interaction Effects}
\change{The contrast in open-ended answers between users who liked or disliked the body}, together with the lack of main effects in confirmatory analyses (in contrast to the moderate to large effects observed in the pilot study) prompted us to dig deeper into the data to look for potential crossover interactions that could be masking main effects. To model SUS and co-presence scores we used cumulative link mixed models from the R package 'Ordinal' (\cite{clmm:2020}) designed to handle repeated measure predictors and ordinal response variables. In each of these models, between-participant variance was specified as a random intercept term. 

\change{\textbf{Presence was well explained by saturated interaction models using condition, session, and sentiment as predictors.}} We took inspiration from the qualitative results; as described in Section \ref{sec:qualdata}, there was a strong divide concerning whether having a virtual body improved the overall telepresence experience, with some participants describing how they appreciated the virtual body while others described how it bothered them. Rather than treat their sentiment towards the virtual body as an outcome variable, we instead tried using it as a predictor to investigate whether sentiment modulated presence effects. Participants' ``yes/no" responses to the question asking whether the virtual body improved the overall experience was inserted into the model as a proxy for their sentiment towards the virtual body. The three-way interaction models predicting SUS and co-presence via condition (SUS: $\beta=6.32$, \textit{CI}$=3.09, 9.54$; co-presence: $\beta=4.82$, \textit{CI}$=2.07, 7.56$), session (SUS: $\beta=3.57$, \textit{CI}$=0.60, 6.54$; co-presence: $\beta=4.48$, \textit{CI}$=1.79, 7.17$), and sentiment (SUS: $\beta=3.34$, \textit{CI}$=0.58, 6.10$; co-presence: $\beta=3.60$, \textit{CI}$=1.19, 6.02$) and their interaction terms were successful (Appendix 2, Interaction Models). Thus, having a virtual body increased senses of PI and co-presence but only when participants held a positive sentiment towards the virtual body, and in a manner that was also sensitive to the participant's current session, with more pronounced differences due to sentiment arising from the approach session than the rat throw session. These saturated interaction models outperformed the null models and additive models (Appendix 2), and the remaining possible unsaturated model configurations. 

\begin{figure*}%
  \centering
        \includegraphics[width=2\columnwidth]{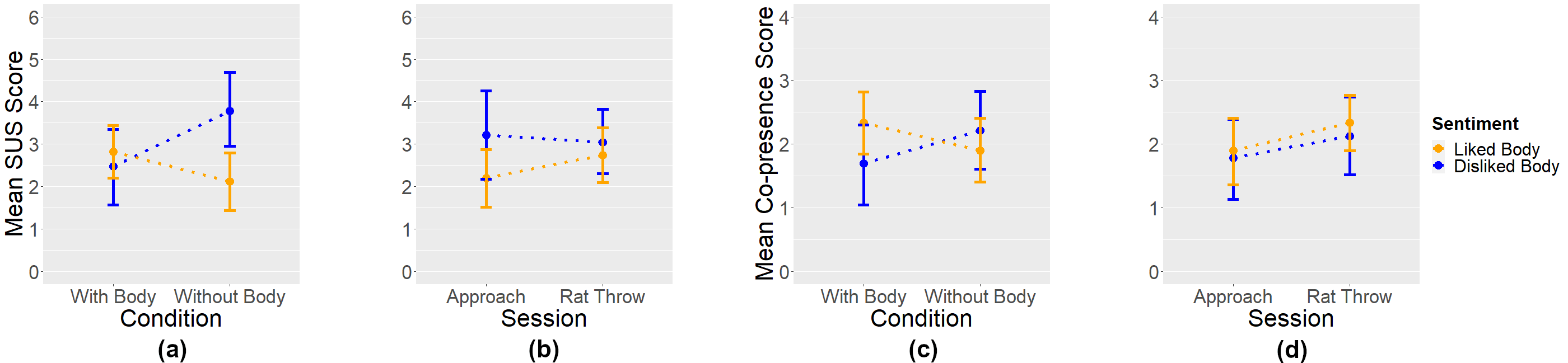}
\caption{Two-way interactions of condition and sentiment in SUS (a) and co-presence (c) and session and sentiment in both metrics, (b) and (d), respectively. Error bars represent 95\% bootstrapped confidence intervals.}
\label{figabc}
\vspace{-0.5cm}
\end{figure*}

\begin{figure}%
  \centering
        \includegraphics[width=\columnwidth]{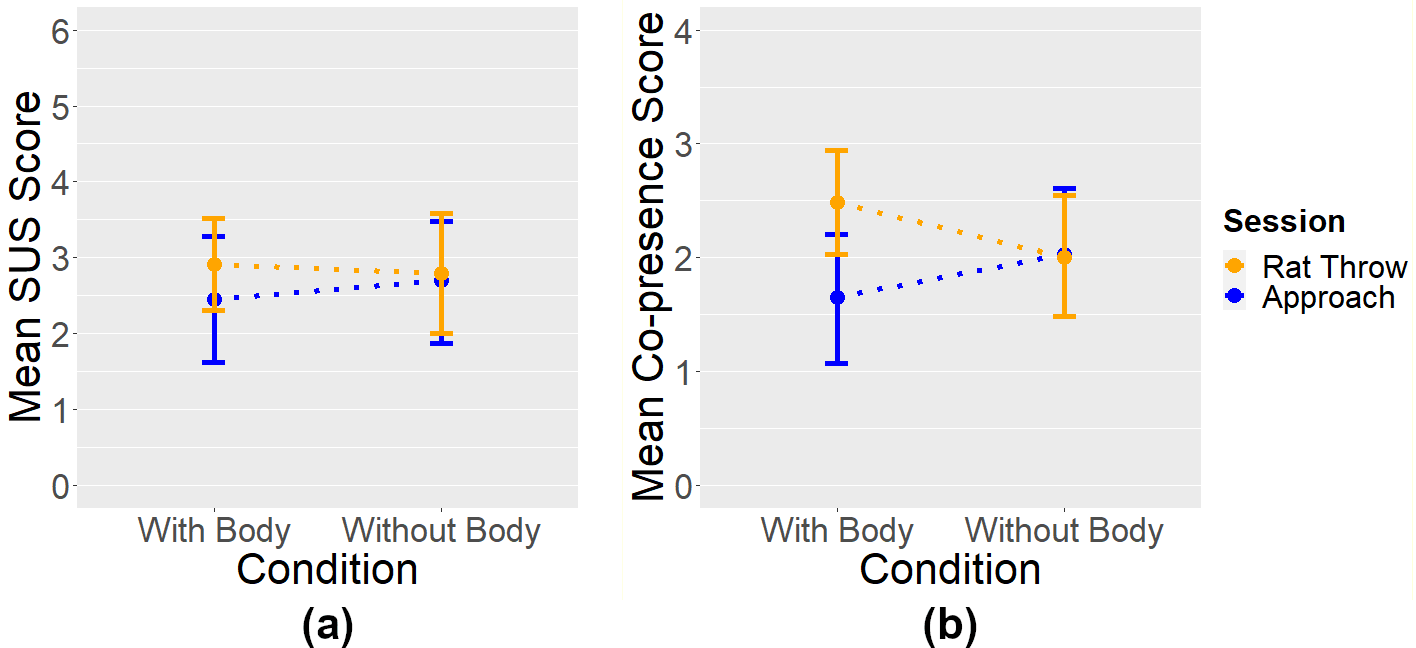}
\caption{Two-way interactions of condition and session in SUS (a) and co-presence (b). Error bars represent 95\% bootstrapped confidence intervals.}
\label{fig:twoway2}
\vspace{-0.5cm}
\end{figure}

\change{\textbf{Sentiment towards the virtual body is key to predicting presence across conditions and sessions.}} Figures \ref{figabc} and \ref{fig:twoway2} display two way interactions between condition and sentiment, session and sentiment, and condition and session, for SUS and co-presence scores. Figure 8a depicts the interaction between condition and sentiment for SUS scores, revealing that scores were similar when participants had a virtual body regardless of whether they liked the virtual body; however, when the virtual body was absent, those who disliked the virtual body recorded \textit{higher} SUS scores than those who liked the virtual body. Figure 8c portrays a similar story for co-presence scores, yet here the larger difference due to sentiment comes during the with body condition rather than in the without body condition. Figure 8b depicts the interaction between session and sentiment for SUS scores, revealing that sentiment towards the virtual body had little effect during the rat throw session, yet during the approach session, those who liked the virtual body recorded higher scores than those who disliked it. This interaction played out differently for co-presence scores (Figure 8d), where generally higher scores were recorded during the rat throw event than in the approach event, and in particular for those who liked the virtual body. Figures 9a and 9b depict the interaction between condition and session for SUS scores and co-presence scores, respectively. Scores tended to be higher in the rat throw condition, and differences due to sentiment, particularly for co-presence scores, were larger in the body condition than in the without body condition.

\begin{figure}%
  \centering
        \includegraphics[width=\columnwidth]{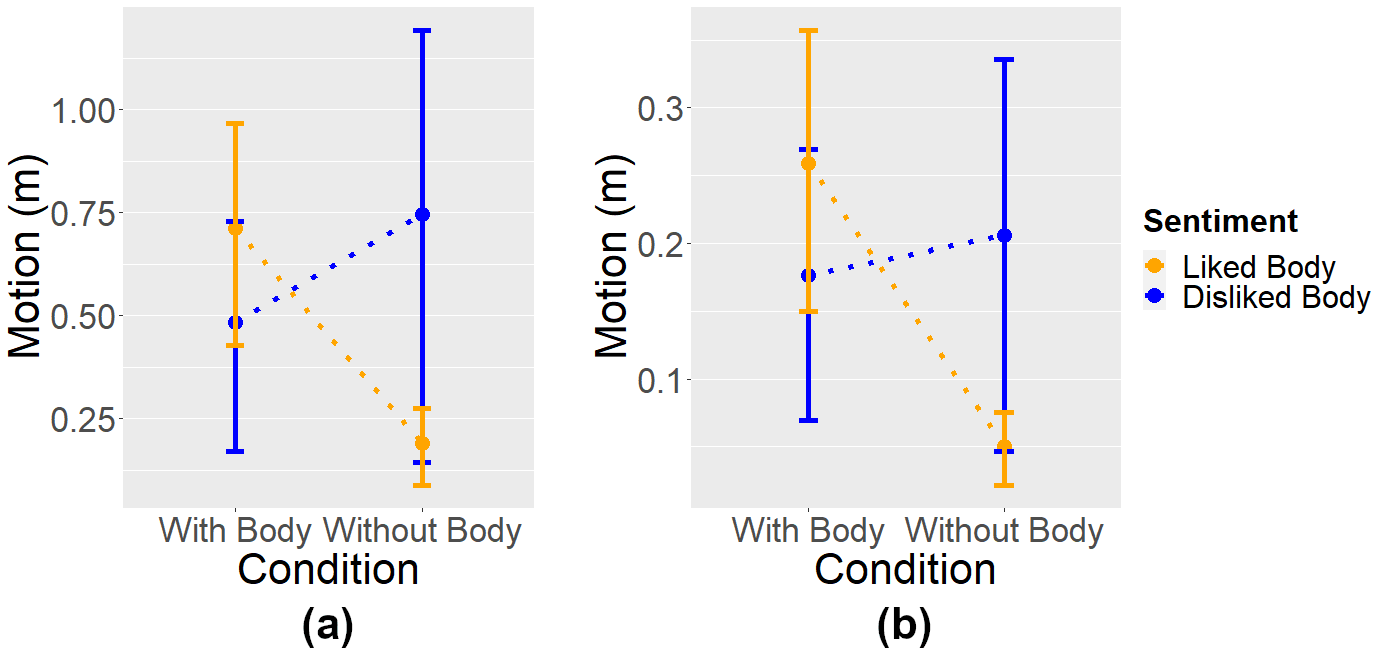}
\caption{Two-way interactions of condition and sentiment measured in hand motions, using either the integration (a) or max vector (b) metric. Error bars represent 95\% bootstrapped confidence intervals.}
\label{fig:handmotion}
\vspace{-0.7cm}
\end{figure}

\change{\textbf{Sentiment additionally predicts the magnitude of motion in response to the rat throw event.}} Although there was some evidence of a main effect of motion per confirmatory analyses, we were curious whether the motion effect would also interact with sentiment. If not, it would suggest that seeing a representation of our hands causes us to move them in response to a threatening stimulus, regardless of whether we like our virtual hands or find their virtual representation to be believable. If instead the motion effect did interact with sentiment, this would suggest that even our implicit, automatic reactions can be shaped by higher level cognition regarding whether we like our virtual hands or find them to be believable. We find evidence in support of the latter. We used the R package 'fitdistrplus' \cite{delignette2015fitdistrplus} to graphically compare fits for our motion distribution, and found it to reconcile best with a gamma distribution. We then fit a gamma regression model using condition, sentiment, and their interaction term to predict integrated hand motion. The interaction term was a good predictor of motion ($\beta=4.60$, \textit{CI}$=1.76, 8.15$), while the main effects of condition ($\beta=-0.73$, \textit{CI}$=0.72, -2.47$) and sentiment ($\beta=-0.62$, \textit{CI}$=0.63, -2.35$) were not. Still, this interaction model significantly outperformed the null model, the condition predictor only model, and the additive model (Appendix 2). Figure \ref{fig:handmotion} depicts this hand motion interaction. For those who did not like the virtual body, movement was largely equivalent whether or not they had a virtual body, while for those who did like the virtual body, they tended to move more when they had it and barely move at all without it.

\subsubsection{Demographic Analyses and Embodiment}
\change{\textbf{Demographics failed to predict presence.}} No demographic information such as age, gender, time spent playing video games, or time spent using VR predicted whether a participant was more likely to respond that having a virtual body improved their overall experience in binomial regression models, nor did they predict differences in SUS or co-presence scores between conditions in ordinal regression models (all model coefficient 95\% bootstrapped \textit{CI}s include 0).

\change{\textbf{Greater embodiment was associated with improved experiences with the virtual body, but not with increased presence.}} The embodiment scores did not predict SUS or co-presence scores in ordinal regression models; however, embodiment scores did moderately predict whether a participant felt that having a virtual body improved the overall experience in a binomial regression model (Appendix 2). Those who felt greater embodiment were more likely to report that the virtual body improved their experience ($\beta=0.88; \textit{CI}=0.28, 1.58$). Despite this relationship, sentiment outperformed the embodiment predictor in interaction models, suggesting that while embodiment could influence sentiment, there remained contributions of variance from sources other than embodiment. Embodiment was thus not included in the final interaction models.
\vspace{-0.1cm}
\change{\subsubsection{Confidence in Reciprocal Viewing}
\textbf{Greater confidence that the host could see the participant's arms was associated with greater presence.} After completing the experiment, participants were asked to rate their confidence that the host could see their hands, separately for each condition, using a 7-point Likert-scale. Participants were significantly more confident that the host could see their hands in the virtual body condition, as revealed by a Wilcoxon signed-rank test (two-sided), \textit{Z}=4.48, \textit{p}$<.001$, \textit{r}=0.57. Furthermore, confidence that the host could see their hands predicted SUS scores ($\beta=0.27; \textit{CI}=0.05, 0.49$) and co-presence scores ($\beta=0.24; \textit{CI}=0.03, 0.46$) in ordinal regression models, although interactions with condition did not reach significance (SUS: $\beta=0.47; \textit{CI}=-0.02, 0.97$; co-presence: $\beta=0.48; \textit{CI}=-0.02, 0.98$). This overall pattern suggests that while participants were more likely to believe that the host could see their movements when they had a virtual body, a stronger belief that the host could see them led to stronger presence regardless of whether they had a virtual body or not.
}


\vspace{-0.1cm}

\section{Discussion}
\label{sec:discuss}

\change{This work suggests that individual differences may play a role in attempts to increase the strength of the PI and feelings of co-presence. The difference between pilot and confirmatory study results, not confirming the pre-registered hypotheses, was initially a surprise since the pilot study hinted towards moderate to large effect sizes for each result. The only consistent results between the pilot and confirmatory analyses were the extra behavioral metrics for PI. However, exploratory analysis on the confirmatory study 
revealed that those who preferred the body experienced greater presence when they had a body, while conversely those who disliked the body experienced greater presence when they \textit{did not} have a body. Thus, the lopsided proportion of people liking the body in the pilot study (18/20) seems to have caused this discrepancy. It is also interesting to note that earlier work on whether seeing arms while watching 360 videos increases presence provided conflicting results, with one paper finding supporting evidence \cite{chen2017effect} while the other did not \cite{danieau2017enabling}. Though the use case here is different, with a truly interactive scenario, further research is merited.}

\change{To delve deeper into the reasons for dependency between sentiment towards the body and feelings of presence, we analyzed the answers to open-ended questions asking why the body did or did not increase the PI.} Thirteen out of 26 participants who did not consider the virtual body increasing PI complained about the \textit{naturalness} of the body. Similar responses came up also in the second forced choice question asking for general preferences (10 out of 23 mentioning \textit{naturalness} as the reason for not preferring the body). \change{This connection is supported by previous research: visual inconsistencies, and inconsistencies in general, seem to have a detrimental effect on PI \cite{garau2008temporal, skarbez2020immersion}, and personalized limbs, on the other hand, increase body ownership (see Section~\ref{sec:related_presence}). \change{The choice of robot arms, based on the use case of real robot arms and avoiding the uncanny valley, seemed to have a stronger effect than we anticipated.} Also, other use cases with virtual agents show that perceived likeness to virtual agents increases acceptance \cite{zhou2014agent}.} Thus, we conclude that there may be many people who benefit from seeing their own arms in immersive telepresence videos, if care is taken to make sure the arms look and feel real. \change{It would be beneficial to repeat the experiment with more realistic appearing arms in the future to confirm this.} 




\change{It is not clear why disliking the body in the approach session caused higher PI scores and having a body in the rat throw caused higher co-presence.} Given the differences in the response-eliciting events between the sessions (see Section \ref{sec:pilot}), we reason that this effect likely arises due to the timescale or the nature of event, or some combination of both factors. Perhaps another human slowly entering and leaving the space where they should have made physical contact with the participant led to a stronger PI break than a briefly thrown object, or led to a longer PI break which a participant could weigh more heavily than a brief one. 

Although our main focus was PI and co-presence, the open-ended data contained many responses that were PSI related. For example, \textit{``it just in some way felt more real cause I could see my hands in that environment, even if they were robot hands"} when stating a preference towards virtual body, or \textit{``because seeing my hands feels like a little bit that everything is unreal"} when stating why PI was stronger without body. Although we used the term \textit{naturalness} when coding responses concerning unnatural virtual arms, these remarks can be also interpreted as concerning the \textit{coherence} of the virtual body {(for example, \textit{``the body felt more being like a pat [sic] of something else then my own.")}}. According to many of these remarks, the computer generated robot arms simply seemed out of place for many participants. Although PI and PSI are generally treated as separate illusions (one can have PSI without PI and vice versa), in our case, having a virtual body that broke PSI seemed to affect PI as well (participants reporting preference towards not having a virtual body reporting significantly lower PI scores). However, this is in line with previous research in the sense that virtual body seems to be an aspect that affects both PI and PSI \cite{slater2010simulating, slater_psi:2009}.
 
This work \change{also} demonstrates the importance of running sufficiently powered studies. If we had trusted the results of our initial 20 participant pilot without performing a confirmatory analysis in a larger sample, we could have falsely concluded that rendering a robot-styled virtual body has a strong positive impact on many facets of the telepresence experience, when the larger, and thus, more likely to be representative, sample indicates that it can have the \textit{opposite} effect depending on how the specific user reacts to the virtual body. \change{To better generalize the results, similar studies with different arms, both high-quality robot arms and personalized human-like arms, should be used; however, both the open-ended answers and Section~\ref{sec:related_presence} hint that better graphics and more individualized arms would increase the presence.} That said, the results concerning the modulating effect of sentiment are experimental, and should be taken with caution until replicated in future experiments.

\vspace{-0.5cm}

\change{\subsection{Limitations}}
The stimuli we chose, throwing the plush rat doll at the participant and violating their personal space, seemed to evoke reactions from the participants, but also to break their presence. The reasons were well documented in the open-ended questions; in the personal space violation, the approach, participants often mentioned that the host came so close it was unrealistic they did not bump into the participant (``\textit{The host essentially passed my arm as she was getting Chewie}" , ``\textit{She came too close, would have touched me, but of course no feeling of it.}"). Similar comments, if not as many, were observed with the rat throwing (``\textit{Being hit by the rat but not really feeling it, causing me to realize it was virtual}" , ``\textit{When i realised that the white mouse is not actually going to touch my skin}"). This risk was already evident in the pilot study, but we chose to keep the same stimuli in the confirmatory study because they did evoke visible reactions in the participants, and we wanted to have an objective measurement of PI besides the questionnaires. \change{We feared that more typical events, such as dropping a pen, would not provoke strong enough reactions from the participants to be registered by our measures. We also note that there may have been individual differences across sessions on how close exactly the host went to the camera, due to inaccuracies in the throwing, but we took steps to make this as consistent as possible.}

Also, always having rat throw in the first session and approach in the second could have caused an HMD accommodation effect (which could also explain the effect condition had in the interactions in Figs.~\ref{figabc}(b,d)
,\ref{fig:twoway2}). However, we argue that this is unlikely; participants took long pauses between sessions to answer questionnaires where the PI could have had time to reset, and even if some effect carried over between sessions, we would expect that more time in the HMD, and thus the second session (approach), would receive higher scores, whereas our results clearly show a trend of higher scores in the first session (rat throw). 

\change{As discussed earlier, the appearance of the robot arms turned out to be a confounding factor. In addition,} participants complained about incorrect visual proportions. These complaints came up especially in the BIP responses (7 occurrences and 4 occurrences without body and with body, respectively). The responses were either referring to participants' sense of own height, the apparent size of the host \change{(for example, \textit{``The host body shape and height seems bit different, may be due to camera"})}, or the relationship of the virtual body to the rest of the environment \change{(for example, \textit{``When I looked around and realized my proportion was out of place")}}. \change{The incorrect proportions were most likely due to the properties of the 360 camera making objects in the video appear larger in comparison to the virtual body. This can be considered a confound causing unintended BIPs in our study.} Finally, there was a risk that using real audio, due to the short physical distance between the host and the participants, could have confounded the study; however, only one participant reported a break in presence due to realizing that the host is in the same room (and two participants even thought they were watching a video), making us confident that real audio did not confound the study.





The main metric used for measuring hand motions during the stimuli (used as the objective measurement of presence), integration, did not provide statistically significant results, whereas both of the robustness measures, max vector and mean vector, did. A downside of using integration is its susceptibility to noise, as 'jittery' participants could potentially artificially inflate their motion values. An upside, however, is the lack of a requirement to choose a baseline, a process which itself could prove to be a source of error if chosen unwisely (though it is worth noting that the max and mean vectors are established metrics when measuring various biosignals and perhaps could be developed into standardized metrics here as well). Ultimately visual inspection of the pilot graphs did not reveal a significant amount of noise, and the motion data in the pilot study were very similar across metrics, which led us to choose integration as the primary metric. Given the consistency of the effect sizes in the same direction among the three, it is likely that they all would have come into agreement with a larger sample and thus more statistical power.


\vspace{0.15cm}
\section{Conclusions and Future work}
Here we reported findings concerning the augmentation of an immersive telepresence experience with a virtual body and its effect on presence and preference. We used questionnaire data and behavioral measures for confirming our hypotheses. We reported the results of our 20 participant pilot study, followed by a 62 participant confirmatory study. \change{The confirmatory study did not replicate the results of the smaller pilot study regarding greater presence and preference with the virtual body, except for a subset of the behavioral metrics.} Exploratory analysis indicated that individual differences could explain the difference between these results. Participants not liking the virtual body often brought up the unnaturalness of the virtual body. During exploratory analysis, we found preference sentiment predicting SUS and co-presence scores; those who liked the virtual body reported higher presence when embodied whereas those who did not reported higher presence scores when not.

There are multiple ways to improve the unnaturalness of the virtual body, \change{a major confounding factor based on open-ended answers}. First, our participants used controllers, even though contemporary hardware can directly track hands, and even fingers. However, the controllers did not come up frequently in the open-ended questions, so it seems this was not a major concern for the participants. Secondly, the visual fidelity of the virtual body could be improved by improving its consistency with the remote environment, for example, by adding environment mapping based reflections \cite{greene1986environment} and lighting that are based on the remote environment; this could potentially reduce the perceived unnaturalness of the body. Additionally, some modern \acp{hmd} have the video pass-through ability, which in theory allows the participants to see their own, actual hands, even though edge artifacts are likely to slightly decrease the quality. Even though one of our motivations was using a robot with real, physical arms, the video passthrough could, in theory, also overlay the participant's real arms on top of the robot's arms. Deficiencies in visual proportions can possibly be improved upon by transforming the \change{virtual body size and virtual interpupillary distance (}IPD\change{)} in software to compensate for the incorrect \change{proportions} caused by the camera since manipulating IPD in VR can be used to scale the perception of sizes and distances \cite{kim2017dwarf, pouke2021plausibility}. 
Finally, we wish to see more research on mixing real and virtual worlds for live telepresence communication. 



\vspace{-0.2cm}
\acknowledgments{This work was supported by the Business Finland project HUMOR 3656/31/2019, Academy of Finland projects PERCEPT 322637, PIXIE 331822, and SRC project COMBAT 293389, and European Research Council project ILLUSIVE 101020977.}

\bibliographystyle{abbrv-doi}

\bibliography{template}
\end{document}